\documentclass[twocolumn,showpacs,preprintnumbers,aps,prc]{revtex4}
\usepackage{epsfig}
\usepackage{dcolumn} 
\usepackage{bm} 
\usepackage{latexsym}
\input {colordvi}
\def\simge{\stackrel{>}{\sim} }

\def\Journal#1#2#3#4{{#1}{\bf #2}, #3 (#4)}

\def\NPA{{Nucl. Phys.}~{\bf A}}
\def\NPB{{Nucl. Phys.}~{\bf B}}
\def\PLB{{Phys. Lett. B}}

\def\PRL{Phys. Rev. Lett.\ }
\def\PRD{{Phys. Rev. D}}
\def\PRC{{Phys. Rev.  C}}
\def\ZPC{{Z. Phys. C}}

\begin{document}
\preprint{BNL-72328}
\title{Acceptance Corrections and Extreme-Independent
Models\\ in Relativistic Heavy Ion Collisions} 
\author{ M.~J.~Tannenbaum}
\altaffiliation{Research supported by U.S. Department of 
Energy, DE-AC02-98CH10886.}
\affiliation{Brookhaven National Laboratory\\ Upton, NY 11973-5000 USA}
\date{\today}
\begin{abstract}
	Kopeliovich's suggestion~\cite{BorisPRC68} to perform nuclear geometry (Glauber) calculations using different cross sections according to the experimental configuration is quite different from the standard practice of the last 20 years and leads to a different nuclear geometry definition for each experiment. The standard procedure for experimentalists is to perform the nuclear geometry calculation using the total inelastic N-N cross section, which results in a common nuclear geometry definition for all experiments. The incomplete acceptance of indivdual experiments is taken into account by correcting the detector response for the probability of measuring zero for an inelastic collision, which can often be determined experimentally.~\cite{E802p0} This clearly separates experimental issues such as different acceptances from theoretical issues which should apply in general to all experiments. Extreme-Independent models are used to illustrate the conditions for which the two methods give consistent or inconsistent results.
	\end{abstract}	
\maketitle

	\section{Introduction}
	In a recent article, Kopeliovich~\cite{BorisPRC68} has discussed the case of nuclear geometry calculations for detectors that are not fully sensitive to the total inelastic nucleon-nucleon (N-N) cross section, i.e. those with limited acceptance. He suggests that the nuclear geometry (Glauber) calculation should be done using different inelastic N-N cross sections according to whether or not the detector is sensitive to inelastic N-N diffractive processes. Explicitly, Kopeliovich suggests to use the N-N non-diffractive cross section, which he estimates to be 30mb, in the nuclear geometry calculation for the PHOBOS and PHENIX detectors, instead of the 42mb total inelastic N-N cross section used by both experiments.~\cite{all4nuclgeom} The general practice by experimentalists is to use the total inelastic N-N cross section in the nuclear geometry calculation and then correct for the limited phase-space coverage (the acceptance) at the detector level~\cite{Brody83, E802-1987, WA80-ZDC}. This practice is based primarily on the fact that the nuclear geometry calculation is then the same for all experiments and so can be easily compared. Secondly, in the case of fixed target experiments, the number of projectile participants, a key nuclear geometry parameter, can be directly measured using a Zero Degree Calorimeter.~\cite{WA80-ZDC,E802PRC44,NA35ZPC52} Thirdly, all experiments have acceptance effects which must be corrected in any case. 
	
	The issue of non-diffractive versus total inelastic N-N cross section is relevant to the definition of projectile participants which can be directly measured using Zero Degree Calorimeters (ZDC). The solid angle of the ZDC is set to a very small forward 
cone around the beam direction $\eta\simge Y^{beam}$, so as to detect only projectile spectators.  The ideal aperture would 
allow the ZDC to measure the full kinetic energy of 
a projectile nucleon in the case of no interaction, and to measure zero energy for 
any nucleon in the projectile that suffered an inelastic collision, including diffraction dissociation. Any inelastic interaction, including diffraction excitation, which causes a projectile nucleon to acquire transverse momentum and lose energy, moves it out of the ideal ZDC aperture.  Thus,  
the energy recorded in the ZDC is proportional to the number of 
non-interacting nucleons (``spectators'') in the projectile, so that the number of nucleons which have interacted (``projectile participants'') is straightforwardly deduced. 

	It is important to note that at RHIC, the ZDC's do not satisfy the ideal criterion and so do not directly measure the number of projectile spectators, hence participants. Nevertheless, it is preferable to keep the nuclear geometry calculations general, in case, for instance, improved ZDC's are installed at RHIC some time in the future. Furthermore, the use of a common nuclear geometry definition by all four RHIC experiments~\cite{all4nuclgeom} has made comparisons among them straightforward and transparent. 

	An additional argument against adjusting the cross section used in the nuclear geometry calculation to account for the experimental configuration is that this procedure does not work in general and may, in certain models, give a different answer from the standard method of including the acceptance in the detector response function. This will be illustrated using examples from extreme-independent models which offer the advantage of explicitly separating the effect of  the nuclear geometry from the detector response. 
  
	\vspace*{0.48in}\section{Extreme-Independent Models} 
	\subsection{Standard procedure to correct for incomplete acceptance}
	  The beauty and utility of having a generally applicable nuclear geometry calculation, rather than one tailored for a particular experiment, is that 
	in the extreme-independent-collision models of B+A nuclear scattering, such 
as the Wounded Nucleon Model~\cite{WNM,Brody83} and Wounded Projectile 
Nucleon Model~\cite{E802-1987,E802p0},   
the effect of the nuclear geometry of the interaction can be calculated 
independently of the dynamics of particle production, which can be taken 
directly from experimental measurements. In these models, the nuclear geometry 
is represented as the relative probability $w_n$ per interaction for 
a given number $n$ of total participants (WNM), projectile participants (WPNM), 
or other basic elements of particle production such as wounded projectile 
quarks (Additive Quark Model--AQM)~\cite{AQM}  or 
binary nucleon-nucleon collisions (NCM), 
integrated over the impact parameter of the B+A reaction. 
Typically, Woods-Saxon densities are used for both the projectile and target 
nuclei, and the nucleon-nucleon inelastic cross section $\sigma^{NN}_{\rm inel}$ appropriate to the 
c.m. energy of the collision is taken. At AGS energies, 30~mb was  
used,~\cite{E802-1987} corresponding to a nucleon-nucleon mean free path of $\sim$2.2 fm at 
nuclear density, while at RHIC, an inelastic N-N cross section of 42~mb is appropriate for $\sqrt{s_{NN}}=200$ GeV.~\cite{egPX200}  Once the nuclear geometry is specified in 
this manner, experimental measurements can be used to derive the distribution 
(in the actual detector) of $E_T$ or multiplicity (or other additive quantity) 
for the elementary collision process, i.e. a wounded nucleon (WNM) or a wounded 
projectile nucleon (WPNM), a wounded projectile quark (AQM) or an N-N collision (NCM), which is then used as the basis of the analysis of a 
nuclear scattering as the result of multiple independent elementary collision 
processes. 

	To illustrate the effect of the detector acceptance, we use the number of collision model (NCM) as  an example to calculate an $E_T$ distribution. The NCM calculation for a B+A reaction is given by the sum:
\begin{equation}
\bigg({d\sigma\over dE_T}\bigg)_{\rm NCM} = \sigma_{BA} \sum^{N_{\rm max}}_{n=1} 
w_n\, P_n(E_T) 
\label{eq:NCM}
\end{equation}
where $\sigma_{BA}$ is the inelastic B+A cross section,    
$w_n$ is the relative probability for $n$ binary collisions in the 
B+A reaction, from 1 to $N_{\rm max}$, and $P_n(E_T)$ is the calculated $E_T$ distribution on the 
detector for $n$ independent N-N collisions.  If $f_1(E_T)$ is the measured $E_T$ 
spectrum on the detector for one detected N-N collision, and $p_0$ is the 
probability for an N-N collision to produce no signal in the detector, then, the correctly normalized $E_T$ distribution for one N-N collision is:
\begin{equation}
P_1(E_T)= (1-p_0)f_1(E_T) +p_0 \delta(E_T) \qquad, 
\label{eq:P1}
\end{equation}
where $\delta(E_T)$ is the Dirac delta function and $\int f_1(E_T)\, dE_T=1$.  
$P_n(E_T)$ (including the $p_0$ effect) is obtained by convoluting 
$P_1(E_T)$ with itself $n-1$ times
\begin{equation}
P_n(E_T) = \sum ^n_{i=0} {{n!} \over {(n-i)!\ i!} } \, 
p_0^{n-i} (1-p_0)^i f_i(E_T) \;\; ,
\label{eq:NCM2}
\end{equation}
where $f_0(E_T)\equiv\delta(E_T)$   
and $f_i(E_T)$ is the $i$-th convolution of $f_1(E_T)$ 
 \begin{equation}
f_i (x)=\int_0^x dy\, f_{1}(y)\,f_{i-1}(x-y)\;\;\; . \label{eq:defcon} 
\end{equation}
Substituting Eq.~\ref{eq:NCM2} into Eq.~\ref{eq:NCM} 
\begin{widetext}
\begin{equation}
\bigg({d\sigma\over dE_T}\bigg)_{\rm NCM} = \sigma_{BA} \sum^{N_{\rm max}}_{n=1} 
w_n\,   \sum ^n_{i=0} {{n!} \over {(n-i)!\ i!} } \, 
p_0^{n-i} (1-p_0)^i f_i(E_T) \qquad ,
\label{eq:NCM2pr}
\end{equation}
\end{widetext}
and 
reversing the indices gives a form that is considerably easier to compute and which is relevant to the present discussion:  
\begin{eqnarray}
&&\bigg({d\sigma\over dE_T}\bigg)_{\rm NCM} = \sigma_{BA} \sum^{N_{\rm max}}_{i=0} 
{w'}_i(p_0)\, f_i(E_T) \nonumber \\
&&= \sigma_{BA} \Big[w'_0(p_0) \delta (E_T) + \sum^{N_{\rm max}}_{i=1} 
{w'}_i(p_0)\, f_i(E_T) \Big] \; , 
\quad  
\label{eq:NCM3}
\end{eqnarray}
where 
\begin{equation}
{w'}_i(p_0) = (1-p_0)^i\, \sum ^{N_{\rm max}}_{n=i} {{n!} \over {(n-i)!\ i!} } \,
p_0^{n-i}\, w_n \qquad ,
\label{eq:NCM4}
\end{equation}
and 
\begin{equation}
{w'}_0(p_0)=\sum ^{N_{\rm max}}_{n=0} p_0^{n}\, w_n = 
\sum ^{N_{\rm max}}_{n=1} p_0^{n}\, w_n 
\label{eq:NCM5}
\end{equation}
is the probability for an inelastic B+A reaction to produce no signal on the detector, where $w_0=0$ by definition.~\cite{BorisPRC68} Thus, the detected cross section for a B+A reaction is $\sigma_{BA}^{det} =\sigma_{BA} (1-w'_0(p_0))$.  

    It is important to emphasize that the acceptance of the experimental measurement can be  accounted for correctly by using the measured N-N cross section in the detector, $\sigma^{NN}_{det}$,  to calculate the probability, $p_0$, that an N-N inelastic collision will produce zero signal on the detector:
    \begin{equation}
    p_0=1- {\sigma^{NN}_{det} \over \sigma^{NN}_{\rm inel}} \qquad ,
    \label{eq:defp0}
    \end{equation}
and then taking $p_0$ into account in the overall detector response.~\cite{E802p0}     
The properly normalized equation for 1 N-N collision on the detector is then given by Eq.~\ref{eq:P1} and the signal for $n$ independent N-N collisions on the detector is given by the binomial distribution Eq.~\ref{eq:NCM2}. Thus, the true mean for $n$ independent $N-N$ collisions on the detector is:
 \begin{equation} 
\langle E_T\rangle^{true}_n=\int E_T\, P_n(E_T)\, dE_T=
n\langle E_T\rangle (1-p_0)\qquad ,
\label{eq:meanETn}
\end{equation}   
which is $n$ times the true mean for 1 N-N collision:
 \begin{equation} 
\langle E_T\rangle^{true}=\int E_T\, P_1(E_T)\, dE_T=
\langle E_T\rangle (1-p_0)\qquad ,
\label{eq:meanET1}
\end{equation}   
where $\langle E_T\rangle$ is the mean of the measured $E_T$ distribution for 1 detected N-N collision, the reference distribution. It is important to contrast Eq.~\ref{eq:meanETn} with the mean of the $n$-th convolution of the observed reference distribution, 
Eq.~\ref{eq:defcon}, 
\begin{equation}
\int E_T\, f_n(E_T)\, dE_T=n \langle E_T\rangle \qquad , 
\label{eq:meanETiobs}
\end{equation}
which is $n$ times the observed $\langle E_T\rangle$, as it should be, and which differs from  
the mean of the distribution $P_n(E_T)$ for $n$ independently interacting projectile nucleons
(Eq.~\ref{eq:meanETn}) by a factor of $1-p_0$ for all $n$.   

     To summarize, the NCM calculation for a B+A reaction is given by Eq.~\ref{eq:NCM}, which uses the relative probabilities, $w_n$, for $n$ independent N-N collisions calculated with the inelastic N-N cross section, $\sigma^{NN}_{\rm inel}$, and takes into account the fraction of inelastic N-N collisions, $p_0$, which produce no signal on the detector, by correcting the response function, $P_n(E_T)$, Eq.~\ref{eq:NCM2}. The result is identical to Eq.~\ref{eq:NCM3}, which is the sum over convolutions of the uncorrected spectrum for $i$ independent detected N-N collisions. The correction for the effect of $p_0$ appears in ${w'}_{i}(p_0)$ (Eq.~\ref{eq:NCM4}), which are thus the relative probabilities for $i$ independent detected N-N collisions.~\cite{refNorm} It is important call attention to a key point of this section: if instead of correcting for the acceptance at the detector level, the correction is made by adjusting the input cross section of the nuclear geometry calculation~\cite{BorisPRC68}, then the ${w'}_{i}(p_0)$  in Eq.~\ref{eq:NCM4} should  correspond to a Glauber calculation with $\sigma^{NN}_{det}=(1-p_0)\sigma^{NN}_{\rm inel}$. 

\subsection{Does using the observed $\sigma^{NN}_{det}$ in the nuclear geometry calculation give the same result?} 
    In the typical static or Glauber Monte Carlo calculation~\cite{Brody83,E802-1987} used to calculate the distribution $w_n$ of binary N-N collisions, a collision is defined when two nucleons pass within a distance $r \leq \sqrt{\sigma^{NN}_{\rm inel}/\pi}$ from each other. For any impact parameter $\vec{b}$ the number of collisions for a given nuclear configuration is calculated and the distribution in the number of collisions is obtained by integrating over all impact parameters. If the distribution $w_n$ is first calculated for a cross section $\sigma^{NN}_{\rm inel}$, then if a smaller cross section, e.g. $\sigma^{NN}_{det}=(1-p_0)\sigma^{NN}_{\rm inel}$, is relevant, the probability for $m$ collisions with cross section $\sigma^{NN}_{det}$,  given $n$ collisions with cross section 
$\sigma^{NN}_{\rm inel}$, is given by the binomial expansion:
\begin{equation}
P(m)|_{n}=  {{n!} \over {(n-m)!\ m!} } \, 
p_0^{n-m} (1-p_0)^m 
\label{eq:Pmn}
\end{equation}
{\em to the extent that the collisions are statistically independent}. The new probability distribution ${w'}_m(p_0)$ for $m$ collisions is obtained by summing over all $n\geq m$, where $0\leq m\leq N_{\rm max}$   
\begin{eqnarray}
 {w'}_m(p_0)&=&\sum^{N_{\rm max}}_{n=m} w_n P(m)|_{n} \cr
            &=& \sum^{N_{\rm max}}_{n=m} w_n  {{n!} \over {(n-m)!\ m!} } \, 
p_0^{n-m} (1-p_0)^m \; , \qquad
\label{eq:wmindep}
\end{eqnarray}          
or, 
\begin{equation}
{w'}_m(p_0)=(1-p_0)^m \sum^{N_{\rm max}}_{n=m} {{n!} \over {(n-m)!\ m!} } \,
p_0^{n-m}\, w_n  \qquad .
\label{eq:wmindep2}
\end{equation} 

	Since the relative probability $w_n$ of $n$ collisions with cross section $\sigma^{NN}_{\rm inel}$ is normalized:
\begin{equation}
\sum^{N_{\rm max}}_{n=1} w_n=1 \qquad ,
\label{eq:wnorm}
\end{equation}
it is easy to see that ${w'}_m$ are also normalized, which follows by reversing the indices $m$ and $n$ in Eq.~\ref{eq:wprnorm} :
\begin{widetext}
\begin{equation}
\sum^{N_{\rm max}}_{m=0}{w'}_m(p_0)=\sum^{N_{\rm max}}_{m=0} (1-p_0)^m \sum^{N_{\rm max}}_{n=m} {{n!} \over {(n-m)!\ m!} } \,
p_0^{n-m}\, w_n  =1 
\label{eq:wprnorm}
\end{equation} 

\begin{equation}
\sum^{N_{\rm max}}_{m=0}{w'}_m(p_0)=\sum^{N_{\rm max}}_{n=1} w_n \sum^{n}_{m=0} {{n!} \over {(n-m)!\ m!} } \,
p_0^{n-m}(1-p_0)^m 
= \sum^{N_{\rm max}}_{n=1} w_n \, (p_0 + (1-p_0))^n =1 \qquad .
\label{eq:wprnorm2}
\end{equation}  
\end{widetext}
	Clearly the ${w'}_m(p_0)$ in Eq.~\ref{eq:wmindep2} and Eq.~\ref{eq:NCM4} are identical.  This proves that where the probability for a collision is proportional to the cross section in question, and when the probabilities for individual inelastic collisions are independent, that the same result for the extreme independent calculation of the $E_T$ distribution in the NCM model for a B+A reaction is obtained whether the detector response per collision is corrected for the probability $p_0$ of recording zero signal for an inelastic collision, or when the actual measured cross section in the detector is used in the calculation of the nuclear geometry. This condition may also apply for the 
AQM, and a specific example of the AQM with the detected cross section used in the nuclear geometry calculation has been given in the literature~\cite{Ochiai} for $\alpha-\alpha$ collisions at the CERN-ISR.~\cite{BCMOR} 

\begin{figure}[htb]


\begin{center}

\includegraphics[scale=0.33,angle=-90]{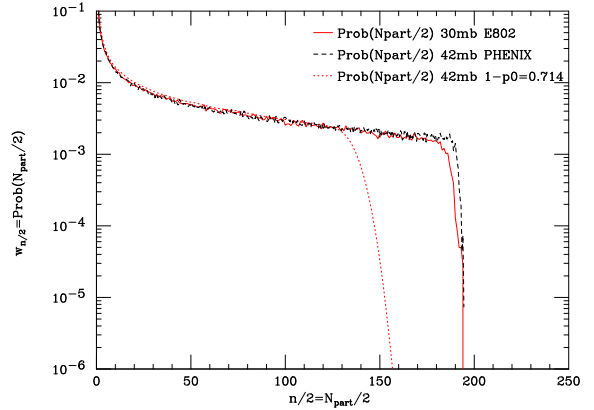}

\caption{$w_{n/2}=$Prob$(N_{part}/2)$  for Au+Au with $\sigma^{NN}_{\rm inel}=30$mb (E802~\cite{E802p0}) or 42 mb (PHENIX~\cite{egPX200}) The dotted line represents ${w'}_{n/2}$ calculated with with $\sigma^{NN}_{\rm inel}=42$mb and $1-p_0=30/42=0.714$. \label{fig:wnm}}

\end{center}\vspace*{-12pt}

\end{figure}

	However, the conditions of statistical independence and linearity of the number of elementary ``collisions" with cross section do not apply for the cases of the Wounded Nucleon Model or the Wounded Projectile Nucleon model. This is shown for the WNM in Fig.~\ref{fig:wnm}, where $w_n$ calculated with either 42mb or 30mb does not change except for the most central collisions, where there is a slight difference---this is rather different from the combinatorial  suppression factor that appears in the correct computation of ${w'}_n(p_0)$ with $1-p_0=30/42=0.714$ (Eqs.~\ref{eq:NCM3},\ref{eq:NCM4}).  
Similar logic applies to the WPNM calculation, where the number of projectile nucleons struck by a target nucleus $A$, with a slightly different p-A cross section, would hardly change for central collisions, and then only for nucleons near the periphery.   

     One may question even for the number of collision model whether the condition for independence of collisions applies. In the Glauber model, an individual nucleon in a projectile nucleus is defined to strike nucleons in the target nucleus when the nucleons in the target are found within a distance $r \leq \sqrt{\sigma^{NN}_{\rm inel}/\pi}$ from the line of the projectile trajectory. In a Glauber Monte Carlo calculation, the randomness of assembling nucleons into a target nucleus with a density 0.16 fm$^{-3}$ (following a Woods-Saxon distribution~\cite{Brody83, E802-1987}) assures the randomness of the number of target nucleons in a cylinder of radius 1.16fm (0.98fm) corresponding to  $\sigma^{NN}_{\rm inel}=42$mb (30mb). Thus the probability to find a nucleon in its spot of phase space in the cylinder (i.e. the probability for a binary collision) for a p+A interaction at a given impact parameter is random so the collisions are independent. However, for a B+A reaction, as the impact parameter varies, the total number of nucleons involved varies in a correlated manner: the more central the interaction, the larger the number of nucleons that can make binary collisions. Thus, the number of binary collisions is not fully random because each of the nucleons participating tends to make  more or fewer binary collisions depending on the impact parameter.  Thus, the distribution in $w_n$ is not binomial for B+A interactions.\footnote{However there are examples in the literature, where only the average number of collisions per incident nucleon is calculated at each impact parameter, and the distribution is {\em assumed} to be binomial or Poisson~\cite{WA80-ZDC,BBR}}  Nevertheless, statistical independence is not unreasonable for the change in probability corresponding to the change in radius of 0.18fm, which is much smaller than the random variation of the number of target nucleons along a given projectile trajectory, so the binomial distribution for the change in the number of collisions with the change in cross section, Eq.~\ref{eq:Pmn}, is probably correct.

     \vspace*{0.36in} \section{Conclusion}	
	We have demonstrated that the use of the experimentally detected cross section in the nuclear geometry (Glauber) calculation is valid in certain classes of models but does not work in general. The procedure which always works~\cite{E802p0} is to perform the nuclear geometry calculation with the inelastic N-N cross section and correct the detector response for the probability of measuring zero for an inelastic collision or other fundamental element of particle production.  


\begin{thebibliography}{99}
\bibitem{refNorm} Note that the normalized conditional probability for $i$ detected collisions, given that at least 1 collision was detected from a B+A reaction is 
${w'}_i(p_0)/(1-{w'}_0(p_0))$ .
\end{thebibliography}
\end{document}